\documentclass[12pt]{article} 
\usepackage{epsfig} 
\date{}

\begin{document} 

\title
{Multiatom cooperative emission  
following single-photon absorption: Dicke-state dynamics}

\author{I E Mazets\footnote{Permanent address: 
A.F. Ioffe Physico-Technical Institute, 194021 
St.Petersburg, Russia} and 
G Kurizki
\\
Chemical Physics Department, Weizmann Institute of Science,
\\ 
76100 Rehovot, Israel
}

\maketitle 

\begin{abstract} 
We investigate conditions under which 
multiatom absorption of a single photon leads to 
cooperative decay. Our analysis reveals the symmetry properties of 
the multiatom Dicke states    underlying the  
cooperative decay dynamics and their spatio-temporal manifestations, 
particularly, the forward-directed  
spontaneous emission investigated by Scully {\em et al}. 
\end{abstract} 


\vspace*{3mm} 


Dicke pioneered  the notion of cooperative spontaneous emission 
by a collection of $N$ atoms, highlighted by 
the ``superradiant'' $N^2$-scaling of the emission rate into 
resonant modes  \cite{dicke}. His work has 
prompted numerous studies of the dependence of cooperative spontaneous 
emission upon the initial preparation (cooperative dipole moment 
and excitation), as well as the spatial symmetry and density (interatomic 
distances) of the multiatom sample 
\cite{cc1,super,sbr,yudson,cc2,molec}. These factors 
determine the degree of cooperativity, which may be attributed to 
multiatom interference of radiated photons. This cooperativity may 
range from maximal 
enhancement (superradiance or superfluorescence), 
first observed in \cite{super}, to 
maximal suppression (subradiance) \cite{sbr,yudson,molec}, 
corresponding to 
constructive or destructive interference, respectively. Similar 
effects in neutron scattering on crystalline lattices 
have also been predicted \cite{kagan}. The most 
powerful systematic means of classifying such effects is the multiatom 
Dicke-states basis, which embodies the symmetry properties of the 
system \cite{cc2,molec}. The 
difficulty is the large multiplicity of Dicke 
states for $N\gg 1$ and their contrasting symmetry properties. The 
cooperative characteristics of the emission may be obscured if many 
Dicke states of different symmetry become entangled or mixed by the 
dynamics of the process. Hence, the recent analysis of Scully {\em et 
al.} \cite {sc1}, implying that $N$ atoms sharing one photon 
absorbed at $t=0$ should subsequently re-emit it {\em only in the 
direction of the absorbed photon} (the forward direction) 
is a nontrivial and perhaps 
counterintuitive manifestation of cooperativity in a large volume 
of {\em randomly distributed} atoms without an initial cooperative 
dipole moment. This aspect is implicit in the far-field analysis
of Rehler and Eberly \cite{cc1, e6x}. 

Here we wish to revisit the foregoing problem using the methods of 
group theory \cite{hm,wig}, so as to elucidate the following important 
questions: Which Dicke states become populated or excited during the 
process? What are their symmetry properties and how are they 
reflected in the spatio-temporal buildup of cooperative spontaneous 
emission? 

Consider the 
absorption of a single photon by a collection of $N$ atoms, initially 
in their internal ground state. 
The photon frequency $ck_0$ is chosen to be exactly at resonance with 
the atomic transition frequency $\omega _0$. We assume that the spatial 
distribution of the atoms is spherically symmetric. For 
the sake of obtaining analytic results, we use a Gaussian probability 
distribution of the random atomic positions  
${\bf r}_j$ inside the sphere: 
\begin{equation} 
{\cal P}({\bf r}) =(\sqrt{\pi }R_0)^{-3}\exp (-r^2/R_0^2). 
\label{P1} 
\end{equation} 
Here $R_0$ stands for the typical size of the spherical sample. 

We expand the wave function of the system ``field + matter'' as follows: 
\begin{equation} 
|\Psi \rangle =\sum _{j=1}^N\alpha _je^{-i\omega _0t}|e_j\rangle + 
\sum _{\bf k} \kappa _{\bf k}e^{-ickt}|{\bf k}\rangle , 
\label{decomp} 
\end{equation} 
where $|e_j\rangle $ denotes all the atoms in the ground state, 
except for the excited $j$th atom, the electromagnetic field 
being in the vacuum state, and $|{\bf k}\rangle $ denotes all the 
atoms in the ground state, with one photon  present in the 
{\bf k} mode. In the present treatment we neglect the effects of photon 
polarization. The Schr\"odinger equation then reads 
\begin{eqnarray} 
i\dot{\alpha }_j&=&\sum _{\bf k}g_{\bf k}e^{i{\bf kr}_j}\kappa _{\bf k}, 
\label{S1a} \\
i\dot{\kappa }_{\bf k}&=&(ck -\omega _0)\kappa _{\bf k} +
\sum _{j=1}^N g_{\bf k}^*e^{-i{\bf kr}_j}\alpha _j  ,   \label{S1b} 
\end{eqnarray} 
where $g_{\bf k}$ is the atom-photon coupling constant for the {\bf k} 
mode. The initial amplitudes in (\ref{decomp}) are expressed by  
\begin{equation} 
\alpha _j(0)=\frac {\exp(i{\bf k}_0{\bf r}_j)}{\sqrt{N}} , \qquad 
\kappa _{\bf k}(0)=0.    \label{ic1} 
\end{equation}
The integration of   (\ref{S1b}) permits us to express 
$\kappa _{\bf k}$ through a time integral involving $\alpha _j$. Then,  
pulling $\alpha _j$ out of the time integral on the assumption that 
$\alpha _j$ decays slowly on the scale of the cooperation time 
(see below), we convert   (\ref{S1a}) into the following equation: 
\begin{equation} 
\dot{\alpha }_j=-\sum _{j^\prime =1}^N \Gamma _{jj^\prime }(t)
\alpha  _{j^\prime } , 
\label{aat} 
\end{equation} 
where 
\begin{equation} 
\Gamma _{jj^\prime }(t)=\sum _{\bf k} |g_{\bf k}|^2 
e^{i{\bf k}({\bf r}_j-{\bf r}_{j^\prime })}\left[ \frac 
{\sin (\omega _0-ck)t}{\omega _0 -ck}-i\frac 
{\cos (\omega _0-ck)t-1}{\omega _0 -ck}\right] 
. \label{complexgamma} 
\end{equation} 
The second term in the square brackets in   (\ref{complexgamma}) 
diverges and thus requires renormalization. It corresponds to the 
Lamb shift of the optical transition, which is, in principle, 
different for each of the $N$ collective excited states 
given by the  mutually orthogonal linear combinations of 
$|e _j\rangle $. This varying part of the Lamb shift 
is of co-operative origin \cite{yudson,cc2}. It can be calculated 
upon noting that the finite sample size and the corresponding wave 
vector spread $\Delta k\sim R_0^{-1}$ remove the ultraviolet 
divergence characteristic of the (unrenormalized)  
Lamb shift for a single atom \cite{mil}. We then find that 
the contribution of co-operative effects to the Lamb shift of a 
collective state is approximately $k_0R_0$ times smaller than the 
decay rate of this state. Hence,  
the {\em variance} of the Lamb shift is insignificant compared to the 
collective decay rate. It allows us in what follows 
to include, as usual, the Lamb shift into the definition of the 
transition frequency $\omega _0$, retaining only the real $\frac 
{\sin (\omega _0-ck)t}{\omega _0 -ck}$ term in  the square brackets 
on the right-hand-side of (\ref{complexgamma}). 

For $t\gg \omega _0^{-1}$ we can then 
substitute in (\ref{complexgamma}) $\frac 
{\sin (\omega _0-ck)t}{\omega _0 -ck}\approx \pi 
\delta ({\omega _0 -ck})$ and the 
expansion ${\bf k}\approx {\bf n}(k_0+\delta k)$, where 
$\delta k\ll k_0$ and ${\bf n}$ is the unit vector in the 
{\bf k} direction, thereby obtaining the following estimation:  
\begin{equation} 
\Gamma _{jj^\prime }(t)
\approx \int \frac{d\Omega _{\bf n}}{4\pi }\, 
\pi |g_{\bf k}|^2\varrho (ck)\vert _{{\bf k}={\bf n}k _0}
e^{ik_0{\bf n}({\bf r}_j-{\bf r}_{j^\prime })}
\Theta [ct -\left| {\bf n}({\bf r}_j-{\bf r}_{j^\prime })\right| ] , 
\label{realgammat} 
\end{equation}
where $\varrho (ck)$ is the density of photon modes, 
$\Theta (x)$ is the Heavyside step function, which is equal to 1 
for $x>0$, $\frac 12$ for $x=0$, and 0 for $x<0$. 

One can see from 
(\ref{realgammat}) that if $ct$ is much smaller than the mean 
interparticle distance then the decay matrix $\Gamma _{jj^\prime }$ 
is {diagonal}, all its diagonal elements being the 
same, equal to the single-atom decay rate 
\begin{equation} 
\gamma _1 =\int \frac {d\Omega _{{\bf n}}}{4\pi }\, 
\pi |g_{\bf k}|^2\varrho (ck)\vert _{{\bf k}={\bf n}k _0} , 
\label{gamma1at} 
\end{equation}
indicating the total absence of cooperativity. 
As the time $t$ increases,  cooperativity is established  
among increasingly more atoms. At 
$t\approx R_0/c$ the collective regime of radiation is 
fully established. In what follows we consider $t\gg R_0/c $. In this     
limit  Eqs.  (\ref{aat}, \ref{complexgamma}) reduce to 
\begin{equation} 
\dot{\alpha  }_j
=-\gamma _1\sum _{j^\prime =1}^N \frac {\sin k_0|{\bf r}_j -
{\bf r}_{j^\prime }|} { k_0|{\bf r}_j -
{\bf r}_{j^\prime }|}\alpha _{j^\prime } , \qquad t\gg R_0/c.
\label{absk} 
\end{equation}

It is extremely difficult to calculate the exact eigenvalues and 
eigenstates of   (\ref{absk}). Therefore, in what follows we introduce 
states that closely approximate the eigenstates of the problem. 
To this end, it is convenient to include 
the phase factors associated with the incident photon momentum into 
the definition of the excited states, 
\begin{equation} 
e^{i{\bf k}_0{\bf r}}|e_j\rangle  \rightarrow |e_j\rangle .    
\label{newst} 
\end{equation}  
The corresponding new probability amplitudes are 
\begin{equation} 
\beta _j\equiv e^{-i{\bf k}_0{\bf r}}\alpha _j.     \label{newpa} 
\end{equation} 
In this basis,   (\ref{absk}) takes the form 
\begin{equation} 
\dot{\beta }_j=-\gamma _1\sum _{j^\prime =1}^N 
F({\bf r}_j-{\bf r}_{j^\prime })\beta _{j^\prime } , 
\label{bett} 
\end{equation}
where 
\begin{equation} 
F({\bf r}_j-{\bf r}_{j^\prime })=
\frac {\sin k_0|{\bf r}_j-
{\bf r}_{j^\prime }|} { k_0|{\bf r}_j -
{\bf r}_{j^\prime }|}e^{-i{\bf k}_0({\bf r}_j -
{\bf r}_{j^\prime })}.             \label{defF}   
\end{equation}

The powerful theory of the permutation group representations and their 
characterization by 
Young tableaus  \cite{hm} provides a general recipe for 
constructing  the functions of arbitrary symmetry with respect to 
permutations. Our case is relatively simple, since we deal with 
a system of two-level atoms. The relevant Young tableaus contain one 
of two rows only, i.e., they are denoted, respectively, by $\{ N \} $ or 
$\{ N -N^\prime , \, N^\prime \} $ with $N-N^\prime \geq N^\prime >0$. 
To construct corresponding wavefunctions, we apply the 
method described in \cite{wig}. 
Consider an operator $\hat {\cal W}$ defined as follows. 
If $\hat {\cal W}$ is applied to the wave function of $N$ 
atoms in their internal ground state, the result is $\sum _{j=1}^N |e_j 
\rangle $. If the $j_1$th, $j_2$th,~...~, $j_m$th  atoms are initially 
excited, then 
\begin{equation} 
\hat {\cal W}|e_{j_1}\rangle |e_{j_2}\rangle \, \dots \, |e_{j_m}\rangle 
=|e_{j_1}\rangle |e_{j_2}\rangle \, \dots \, |e_{j_m}\rangle 
{\sum _j}^\prime |e_j\rangle   , 
\label{spr} 
\end{equation}  
where ${\sum _j}^\prime $ denotes the sum over $j\neq j_1,\, 
j_2, \, \dots \, , \, j_m$. This $\hat {\cal W}$ commutes with any 
product of {\em generalized permutation operators}, which 
interchange not only the internal-state atomic variables, but 
the coordinate-dependent phase factors as well: 
\begin{equation} 
\hat{\cal O }_{jl}|e_i\rangle =
\left \{ \begin{array}{ll} 
|e_l\rangle , & i=j \\
|e_j\rangle , & i=l \\
|e_i\rangle , & i\neq j,l \end{array} \right. 
.\label{gpo} 
\end{equation}
Thus $\hat {\cal W}$ conserves the symmetry type of the 
state. Upon applying $\hat{\cal W}$ to the wave function of $N$ atoms 
in the ground state, which is totally symmetric with respect 
to generalized permutations, we obtain  
the totally symmetric state  with one atom excited  
\begin{equation} 
|\phi ^{\{ N \} }\rangle =\frac 1{\sqrt{N}}\sum _{j=1}^N|e_j\rangle ,
\label{YN}
\end{equation}   
which constitutes a one-dimensional group representation 
characterized by the Young tableau $\{ N\} $. 
We now construct from $N$ linearly independent states $|e_j\rangle $, 
by orthogonalization to $|\phi ^{\{ N\} \} }\rangle $,  
the $N-1$ states  
\begin{eqnarray} 
|\phi ^{\{ N-1,1\} }_l\rangle &=&
-\frac 1{\sqrt{N}}|e_N\rangle +\sum _{j=1}^{N-1} 
\left[ \frac {1+(1/\sqrt{N})}{N-1}-\delta _{jl}\right] |e_j\rangle 
\nonumber \\ && 
\equiv \sum _{j=1}^Nf^l_j|e_j\rangle , 
\qquad  l=1,\, 2,\, \dots , \, N-1, 
\label{YN-11}
\end{eqnarray} 
which comprise the basis of the irreducible representation 
characterized by the Young tableau $\{ N-1, \, 1\} $. The states 
(\ref{YN-11}) 
are normalized to 1 and orthogonal to each other and to $|\phi ^{\{ N\} }
\rangle $. Any product 
of pairwise operators (\ref{gpo}) transforms any of the wave functions 
(\ref{YN-11}) into a linear combination of these functions, without 
adding terms containing $|\phi ^{\{ N \} }\rangle $. 
This procedure may be extended to the construction of doubly-excited 
states with the Young tableau $\{ N-2,\, 2\} $ and so on. 
 
Expanding the wave function of the singly-excited atomic states as 
\begin{equation} 
|\psi _{exc}\rangle =c^{\{N\} }|\phi ^{\{ N\} }\rangle + 
\sum _{l=1}^{N-1}c^{\{ N-1,1\} }_l|\phi ^{\{ N-1,1\} }_l\rangle , 
\label{psiexc}
\end{equation} 
we arrive at the following set of equations, whose terms 
are explained below: 
\begin{eqnarray} 
\dot{c}^{\{ N\} }&=&-\gamma _{col}{c}^{\{ N\} }-
\sum _{l=1}^{N-1}s_l^*c^{\{ N-1,1\} }_l   ,\label{S2a} \\
\dot{c}^{\{ N-1,1\} }_l&=&-s_lc^{\{ N\} }- \sum _{l^\prime =1}^{N-1} 
Q_{ll^\prime } {c}^{\{ N-1,1\} }_{l^\prime }, \label{S2b} 
\end{eqnarray} 
satisfying the initial conditions 
\begin{equation} 
\dot{c}^{\{ N\} }(0)=1, \qquad c^{\{ N-1,1\} }_l(0)=0, \quad   
l=1,\, 2, \, \dots  \, N-1. \label{ic2} 
\end{equation}  
The {\em collective decay rate of the fully symmetric state} is 
then found to be [cf. (\ref{defF})]
\begin{equation} 
\gamma _{col}=\frac{\gamma _1}N\sum _{j=1}^N\sum _{j^\prime =1}^N 
F({\bf r}_j-{\bf r}_{j^\prime }).   
\label{colg} 
\end{equation} 
The coupling term in   (\ref{S2a}), mixing the fully symmetric state and 
the $l$th state of lower symmetry, appears because the states 
(\ref{YN}), (\ref{YN-11}) {\em are not the exact eigenstates} of the decay 
operator on the right-hand-side of   (\ref{bett}). However, they 
provide a good approximation thereof, because the mixing 
is characterized by the coupling strength  
\begin{equation} 
s_l=\frac{\gamma _1}{\sqrt{N}}\sum _{j=1}^N\sum _{j^\prime =1}^N 
f^l_j F({\bf r}_j-{\bf r}_{j^\prime }), 
\label{coupl} 
\end{equation} 
and is weak (see below) compared to the collective decay rate. 
The coefficients
\begin{equation} 
Q_{ll^\prime }={\gamma _1}\sum _{j=1}^N\sum _{j^\prime =1}^N 
f^l_j f^{l^\prime }_{j^\prime }F({\bf r}_j-{\bf r}_{j^\prime })   
\label{Qll12} 
\end{equation}
in   (\ref{S2b}) describe the decay of the $l$th state of the 
$\{ N-1,1 \} $ symmetry type if $l=l^\prime $, or the mixing  
of states with $l\neq l^\prime $. 
The level scheme, the coupling and decay channels described by 
(\ref{S2a},~\ref{S2b}) are shown in Figure 1. 

\begin{figure} 
\begin{center} 
\centerline{\epsfig{file=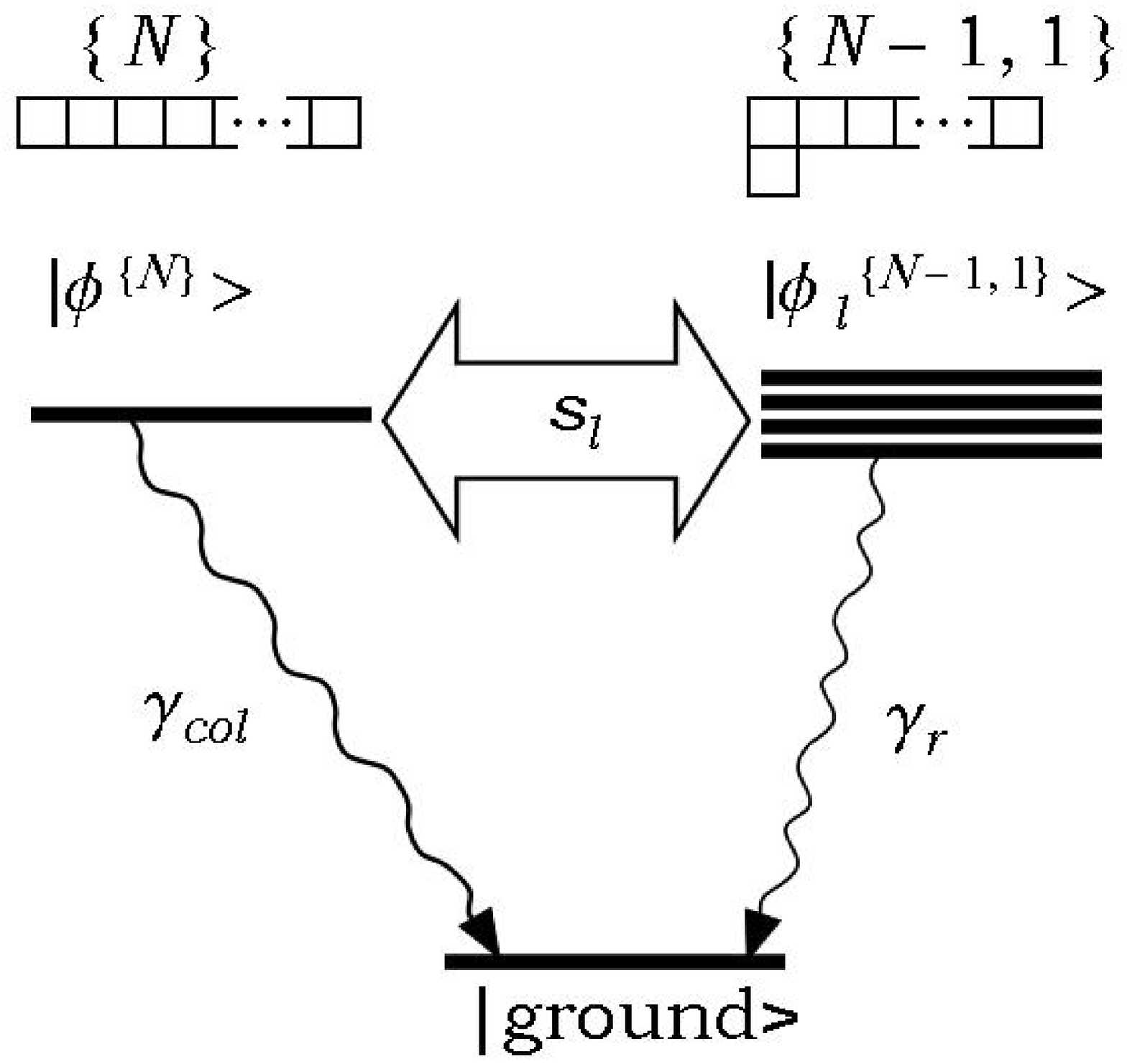,width=5.5cm}}
\end{center} 

\begin{caption} 
{Schematic representation of the processes described by 
  (\ref{YN}, \ref{YN-11}). Young tableaus characterizing 
symmetry of the satates are indicated.}
\end{caption} 
\end{figure} 

Let us first calculate the collective decay rate $\gamma _{col}$. 
Since $N\gg 1$, performing the double sum in   (\ref{colg}) 
and dividing it by $N^2$ is equivalent to  averaging over the atomic 
positions ${\bf r}_j,\, {\bf r}_{j^\prime }$:   
\begin{equation} 
\gamma _{col}={\gamma _1}N\int d^3{\bf r}_j\int d^3{\bf r}_{j^\prime }
\, {\cal P}({\bf r}_j)  {\cal P}({\bf r}_{j^\prime }) 
F({\bf r}_j-{\bf r}_{j^\prime }).  
\label{cg-av1} 
\end{equation}
Here we assume that there is no correlation between the positions 
of different atoms, which is true for gas atoms or dopants in 
a crystal, so that the 
two-particle probability distribution reduces to a product of 
single-particle distribution functions. To evaluate the integral in 
  (\ref{cg-av1}), we recall that  
\begin{equation} 
\frac {\sin k_0|{\bf r}_j-
{\bf r}_{j^\prime }|} { k_0|{\bf r}_j -
{\bf r}_{j^\prime }|}=\int \frac {d\Omega _{\bf n}}{4\pi }
e^{-i{k}_0{\bf n}({\bf r}_j -{\bf r}_{j^\prime })},  
\label{n1}
\end{equation} 
where {\bf n} is a unit vector whose direction is uniformly 
distributed over sphere. Then the integrals of the Gaussian type 
are readily evaluated: assuming that the sample size 
is much larger than the resonant wavelength,  
\begin{equation} 
k_0R_0\gg 1,    \label{bigsample} 
\end{equation} 
we arrive at 
\begin{equation} 
\gamma _{col}=\gamma _1 N (k_0R_0)^{-2}  .    \label{res1} 
\end{equation} 
The factor $(k_0R_0)^{-2}$ is the effective 
solid angle of the collective forward emission of a photon. 

This collective process prevails over the incoherent scattering by 
individual atoms if $N(k_0R_0)^{-2}\gg 1$. Since $N \sim \eta R_0^3$, 
$\eta $ being the atomic density, this condition is equivalent 
to the requirement to have a large number of atoms in a cylinder  
whose length is equal to the sample size in the incident-photon 
direction and the cross-section area of is the order of the wavelength 
squared, i.e. $\eta R_0k_0^{-2} \gg 1$. 

The states of the $\{ N-1,1 \} $ symmetry become  
populated as well. To the first order of the perturbative analysis, 
\begin{equation} 
c^{\{ N-1,1 \} }_l(t)=-s_l\int _0^tdt^\prime \, c^{ \{ N\} }(t^\prime )
=-\frac {s_l}{\gamma _{col}}\left( 1-e^{-\gamma _{col}t} \right)  . 
\label{pt1} 
\end{equation} 
To perform the calculations with an accuracy  
$\sim 1/\sqrt{N}$, it is sufficient to make in   (\ref{coupl}) 
the  approximation 
\begin{equation} 
f^l_j\approx N^{-1}-\delta _{jl}    .   \label{approxf}
\end{equation}
Then 
\begin{eqnarray} 
s_l&=&-\gamma _1\sqrt{N}\left[ \frac 1N\sum _{j=1}^N F({\bf r}_l-
{\bf r}_j)-\frac 1{N^2}\sum _{l=1}^N\sum _{j=1}^N F({\bf r}_l-
{\bf r}_j)\right] \nonumber \\ 
&=&-\gamma _1\sqrt{N}\left[ \int d^3{\bf r}_j\, {\cal P}({\bf r}_j) 
F({\bf r}_l-{\bf r}_j)- \right. \nonumber \\ && 
\left. \int d^3{\bf r}_l \int d^3{\bf r}_j\,
{\cal P}({\bf r}_l) {\cal P}({\bf r}_j) F({\bf r}_l-{\bf r}_j)\right] 
.\label{slgtv} 
\end{eqnarray} 
Straightforward but lengthy calculations using   (\ref{n1}) yield 
in the limit of   (\ref{bigsample}): 
\begin{equation}
c^{\{ N-1,1 \} }_l(t)=\frac {2i{\bf k}_0{\bf r}_l}{\sqrt{N}(k_0R_0)^2} 
\left( 1-e^{-\gamma _{col}t} \right)  . 
\label{pt2} 
\end{equation}

If within the time interval $t\gg \gamma _{col}^{-1}$ we 
{\em do not detect} 
an emitted photon by a perfect detector with 100 \% counting 
efficiency, this implies that the atoms have been coherently transferred 
to the state 
\begin{equation} 
|\psi _r\rangle = {\cal A}\sum _{l=1}^{N-1}i{\bf k}_0{\bf r}_l 
|\phi ^{ \{ N-1,1\} }_l\rangle   .   \label{rsct}  
\end{equation}
Its normalization coefficient can be found in the limit $N\gg 1$ to be 
\begin{equation} 
{\cal A}=\frac 1{\left[ \sum _{l=1}^{N-1} ({\bf k}_0{\bf r}_l)^2
\right] ^{1/2}}
\approx \frac 1{\left[ N\int d^3{\bf r} {\cal P}({\bf r}) 
({\bf k}_0{\bf r})^2 \right] ^{1/2}} =\sqrt{ \frac 2{Nk_0^2R_0^2}}. 
\label{Arsc} 
\end{equation}  
From   (\ref{pt2}) we can calculate the probability of such an 
outcome, namely, that no photon is scattered forward during 
the decay of the fully symmetric state and, instead, 
the new state (\ref{rsct}) is formed and then decays. The probability 
appears to be small, $\sim (k_0R_0)^{-2}\ll 1$, 
that is of the order of the body angle characteristic for 
the superradiant forward scattering. Yet for 
{\em mesoscopic} samples it may be non-negligible, as argued below.

Now we may find the explicit coefficients of the expansion 
$|\psi _r\rangle =\sum _{j=1}^N h_j |e_j\rangle $ for   (\ref{rsct}). 
To the accuracy $\sim 1/\sqrt{N}$ we find that 
\begin{equation} 
h_j=i{\cal A}\left( {\bf k}_0{\bf r}_j -\frac 1N \sum _{j^\prime =1}^N 
{\bf k}_0{\bf r}_{j^\prime }\right) . 
\label{h1} 
\end{equation} 
The first term in the brackets  is of the order of $k_0R_0$, 
while the second term is of the order of $k_0R_0/\sqrt{N}$. We can 
therefore set $h_j\approx i{\cal A}{\bf k}_0{\bf r}_j$. 
The radiative decay rate of the state (\ref{rsct}) is then  
\begin{eqnarray} 
\gamma _r &=&\gamma _1\sum _{j=1}^N\sum _{j^\prime =1}^N 
h_jh_{j^\prime } F({\bf r}_j-{\bf r}_{j^\prime }) \nonumber \\ 
&=&\frac {2\gamma _1}{Nk_0^2R_0^2} \sum _{j=1}^N\sum _{j^\prime =1}^N 
({\bf k}_0{\bf r}_j)({\bf k}_0{\bf r}_{j^\prime }) 
F({\bf r}_j-{\bf r}_{j^\prime }) \nonumber \\ 
&=&\frac {2N\gamma _1}{k_0^2R_0^2} \int d^3{\bf r}_{j}
\int d^3{\bf r}_{j^\prime }\, {\cal P}({\bf r}_{j})
{\cal P}({\bf r}_{j^\prime })
({\bf k}_0{\bf r}_j)({\bf k}_0{\bf r}_{j^\prime }) 
F({\bf r}_j-{\bf r}_{j^\prime }).   \label{grscat} 
\end{eqnarray}
Finally, we obtain 
\begin{equation} 
\gamma _r = \frac {\gamma _1N}{2k_0^4R_0^4} .  
\end{equation} 

The foregoing analysis has shown that the probability of the phonon 
emision first decreases as $\exp (-2\gamma _{col}t)$. For $t\gg 
\gamma _{col}^{-1}$, an ``afterglow'' due to photon reabsorption into 
the states of $\{ N-1,\, 1\} $ symmetry may occur with probability 
$\sim (k_0R_0)^{-2}\exp (-2\gamma _rt)$. The entire emission process 
occurs into the forward preferred direction, consistently with the 
results of \cite{sc1}. 

Since the cross-section of resonant photon absorption is $\sim
k_0^{-2}$,  and the atomic number density is $\sim N/R_0^3$, the
forward-emission  enhancement factor $N/(k_0R_0)^2$ is of the order of
the optical  density of the sample. Can this enhancement, 
$N/(k_0R_0)^2\gg 1$, be consistent  with uniform excitation
probability over the sample [cf. Eq.~(\ref{YN})]?  It can, e.g., if we
use an auxiliary strong laser field acting on a different transition
to cause the the Autler-Townes splitting  of the optical transition at
$\omega_0$, thereby reducing the optical density of the sample at
$\omega _0$ well below 1. After a low-probability non-coincidence
event (detection of the  signal photon only) that signifies the
absorption of the probe photon in the sample, the auxiliary field
should be rapidly switched off, thus restoring the large  optical
density and allowing the Dicke-state dynamics described above. The
need to switch off the Autler-Townes splitting at the proper time to
observe the enhancement of the single-photon  emission stresses the
relevance of the statement \cite{sc1} that ``timing is everything''.

 Our analysis has provided new, more detailed insights into the
buildup of the coperativity in space and time in the spontaneous
emission process triggered by the controlled 
 absorption of a single
photon \cite{drob} at $t=0$.  It has underscored the dominance {\em of
the symmetric Dicke state}, but  only in the long-time and
large-sample asymptotic regime.   Other cooperative states,  of
lower symmetry, become mixed with the symmetric state as the decay 
process unfolds and add slower emission rate in the forward direction.
However, their contribution is non-negligible only if the sample is
 mesoscopic, say $k_0R_0\, ^<_\sim \, 10$. This analysis thus 
corroborates the results of Scully {\em et al.} \cite{sc1}  in the
large-sample limit.
 
 This work is supported by the
German-Israeli Foundation, the EC  (SCALA IP), and the
ISF. I.E.M. acknowledges also support   from the program Russian
Leading Scientific Schools (grant 9879.2006.2).

\newpage 
 
\end{document}